\documentstyle[epsf,aas2pp4]{article}

\newcommand{\etal}{{et al.}\hspace{1mm}}

\lefthead{Bryan \& Norman}
\righthead{Statistical Properties of X-ray Clusters}

\begin{document}


\title{Statistical Properties of X-ray Clusters: Analytic and
Numerical Comparisons} 

\author{Greg L. Bryan}
\affil{Physics Department, Massachusetts Institute of Technology, 
Cambridge, MA 02139}

\and

\author{Michael L. Norman\altaffilmark{1}}
\affil{Astronomy Department, University of Illinois at Urbana-Champaign, 
Urbana, IL 61801}

\altaffiltext{1}{National Center for Supercomputing Applications,
405 North Mathews Avenue, Urbana, IL 61801}


\begin{abstract}
We compare the results of Eulerian hydrodynamic simulations of cluster
formation against virial scaling relations between four bulk
quantities: the cluster mass, the dark matter velocity dispersion, the
gas temperature and the cluster luminosity.  The
comparison is made for a large number of clusters at a range of
redshifts in three different cosmological models (CHDM, CDM and OCDM).
We find that the analytic formulae provide a good description of the
relations between three of the four numerical quantities.  The fourth
(luminosity) also agrees once we introduce a procedure to correct for
the fixed numerical resolution.  We also compute the normalizations
for the virial relations and compare extensively to the existing
literature, finding remarkably good agreement.
The Press-Schechter prescription is calibrated with the
simulations, again finding results consistent with other authors.  We
also examine related issues such as the size of the scatter in the virial
relations, the effect of metallicity with a fixed pass-band, and the
structure of the halos.  All of this is done in order to establish a
firm groundwork for the use of clusters as cosmological probes.
Implications for the models are briefly discussed. 
\end{abstract}

\keywords{galaxies: clusters, X-rays: galaxies, methods: numerical}

\twocolumn


\section{Introduction}

The statistics of X-ray clusters can serve as an excellent probe of
cosmology.  The luminosity function of clusters is defined in a
straightforward manner both observationally (\cite{hen92};
\cite{ebe95}; \cite{bur96}) and numerically (\cite{cen92};
\cite{chdm_xray}), although as we shall see this is a computationally
difficult task.  The temperature function is similarly obtainable 
(\cite{hen91}; \cite{dav93}), however, here it is the observational
data which are more challenging to obtain.  

One uncertainty in performing this comparison is the possibility of
systematic errors in the theoretically derived cluster temperatures
and luminosities.  This has been investigated in a number of ways,
primarily by testing individual methods developed against simplified
problems with known solutions.  So, the Lagrangian smoothed particle
hydrodynamics (SPH) method combined with a P$^3$M code for computing
the gravitational interaction has been examined by Evrard
(1988)\markcite{evr88}, Couchman, Thomas \& Pearce
(1995)\markcite{cou95} and others; while SPH with a gravitational tree
code has been tested in a number of papers (\cite{her89};
\cite{kat96};
\cite{nav93}).  A novel modification of SPH was described by Shapiro
\etal (1996)\markcite{sha96}, along with a number of comparisons
against known results.  For Eulerian codes, method papers with some
tests include: Cen (1992)\markcite{cen92} for a first-order grid based
code; Ryu \etal (1993)\markcite{ryu93} for a total variation
diminishing method; Anninos, Norman \& Clarke (1994)\markcite{ann94}
for a two-level nested grid scheme, and Bryan \etal
(1995)\markcite{PPM} for a code based on the piecewise parabolic
method.

However, a number of recent papers have taken up the question of
accuracy and consistency with a cosmological context as their primary
focus.  For Eulerian codes (with fixed comoving spatial resolution),
this issue has been addressed by Anninos \& Norman
(1996)\markcite{ann96} (hereafter AN96) who performed simulations with
a two level hierarchical method and simulated the formation of a
moderately rich cluster with five different resolutions ranging from
1600 to 100 kpc per cell.  They found that although the temperature,
mass and velocity dispersion of a cluster were reasonably well
determined with the lower-resolution simulations, the total luminosity
had not converged even for the highest resolution run.  The bolometric
luminosity (for this single cluster) behaved as
\begin{equation}
L_{tot} \propto \frac{1}{\Delta r^{\alpha}},
\label{eq:lumin_resolution}
\end{equation}
where $\Delta r$ is the spatial resolution of the simulation and
$\alpha = 1.17$.

Another way to check the results of simulations (and the route taken
in this paper) is to compare them against the predictions of
approximate analytic models.  Although agreement does not guarantee
correctness (as both methods are only approximations to the full
solution), concordance would increase our confidence in both methods.
Also, simple analytic arguments may only identify a scaling property
between quantities without specifying a normalization, which can
however, be fixed by numerical simulation (or by further
assumptions). This describes the scaling laws that come from
considering clusters as spherical clouds of gas in hydrostatic
equilibrium.  Navarro, Frenk \& White (1995)\markcite{NFW}
(hereafter NFW) recently compared the results of six clusters
simulated with SPH in a Cold Dark Matter (CDM) scenario against these
scaling relations (at $z=0$).  They find good agreement over a wide
range of luminosity, mass and temperature, but claim that clusters
from Eulerian simulations (such as in \cite{kan94a} and
\cite{cdm_xray}) do not.

Another analytic method is that initially described by Press \&
Schechter (1974)\markcite{pre74} which predicts the mass distribution
of collapsed objects.  There have been a number of comparisons between
its predictions and the results of N-body simulations (\cite{efs88};
\cite{bon91}; \cite{lac96}).  Using the scaling results, this theory
can be extended to produce the temperature (\cite{eke96}) and
luminosity distribution functions.  Since this is one of the most
widely used constraints on the amplitude of mass fluctuations, it is
important to check its validity.

In this paper, we make a detailed comparison between simulation
results and the adiabatic scaling laws as well as the Press-Schechter
formalism with extensions.  This allows us to gauge the accuracy and
consistency of both methods, leading to firmer conclusions regarding the
viability of the cosmology modelled.  The paper is laid
out as follows.  In section~\ref{sec:scaling}, we review the scaling
relations, including a modification to take into account the finite
resolution of Eulerian codes.  We then compare these to the results of
CDM and Cold plus Hot Dark Matter (CHDM) simulations at a variety of
redshifts.  In section~\ref{sec:distribution}, we examine the mass,
temperature and luminosity distribution functions, including the
effects of finite band-pass and line emission.  These are compared
against the Press-Schechter plus scaling theory (extended to include
the additional complications in the luminosity function).  In
section~\ref{sec:profiles} we briefly examine the profiles of temperature and
velocity dispersion to assess the accuracy of the isothermal
models assumed in extending the Press-Schechter work.  Finally, in
section~\ref{sec:chap6_conclusion}, we discuss our results and comment
on the viability of the models simulated.


\section{Scaling Relations}
\label{sec:scaling}

Here we review the scaling relations between cluster bulk properties
through the assumption of a pressure supported isothermal sphere for
both the gas temperature $T$ and one-dimensional collisionless
velocity dispersion $\sigma$ of the dark matter particles.  The
assumption of a specific density profile (here the isothermal sphere)
is not required to obtain the scaling behaviour, but is needed to
determine the constant of proportionality between the given quantities. 

These relations were used by Kaiser (1986)\markcite{kai86} to describe
the evolution of `characteristic' quantities, largely driven by the
non-linear mass ($M_{nl}$), defined via equation~(\ref{eq:sigma})
below, as well as to derive relations between distribution functions
at different epochs.  We do not explicitly test these because they are
uniquely specified by the non-linear mass (which we do examine) and
the scaling relations discussed below.


\subsection{Scaling review and normalization}
\label{sec:scaling_review}

In the isothermal distribution function, the density is related
to the velocity dispersion (\cite{bin87}):
\begin{equation}
\rho(r) = \frac{\sigma^2}{2 \pi G r^2}.
\label{eq:rho_sigma}
\end{equation}
If we define $r_{vir}$ as the radius of a spherical volume
within which the mean density is $\Delta_c$ times the critical density
at that redshift ($M = 4\pi r_{vir}^3 \rho_{crit} \Delta_c/3$),
then there is a relation between the virial mass and the one-dimensional 
velocity dispersion:
\begin{eqnarray}
\label{eq:mass_sigma_a}
\sigma & = & M^{1/3} \left(H^2(z) \Delta_c G^2/16\right)^{1/6} \\
\label{eq:mass_sigma}
       & = & 476 f_\sigma
	     \left(\frac{M}{10^{15} M_\odot}\right)^{1/3}
	     \left(h^2 \Delta_c E^2\right)^{1/6} \hbox{km/s}
\end{eqnarray}
In the second line we have introduced a factor $f_\sigma$ which will
be used to match the normalization from the simulations.
The redshift-dependent Hubble constant can be
written as $H(z) = 100$ $h$ $E(z)$ km/s/Mpc with the function $E^2
= \Omega_0 (1+z)^3 + \Omega_R (1+z)^2 + \Omega_\Lambda$ dependent on
three contributions:
\begin{equation}
\Omega_0 = \frac{8 \pi G \rho_0}{3H_0^2}; \quad
\Omega_R = \frac{1}{(H_0 R)^2}; \quad
\Omega_\Lambda = \frac{\Lambda}{3 H_0^2}.
\end{equation}
Here, $\rho_0$ is the non-relativistic matter density, $R$ is the
radius of curvature and $\Lambda$ is the cosmological constant.

The value of $\Delta_c$ is taken from the solution to the
collapse of a spherical top-hot perturbation under the assumption that
the cluster has just virialized (\cite{pee80}).  Its
value is $18\pi^2$ for a critical universe but has a dependence on
cosmology through the parameter $\Omega(z) = \Omega_0 (1+z)^3 / E(z)^2$.
We have calculated this for the cases $\Omega_\Lambda = 0$
(\cite{lac93}) and $\Omega_R = 0$ (\cite{eke96}), fitting the results
with:
\begin{eqnarray}
\Delta_c = 18\pi^2 + 82x - 39x^2
	\quad \hbox{for} \quad \Omega_R = 0 \\
\Delta_c = 18\pi^2 + 60x - 32x^2
	\quad \hbox{for} \quad \Omega_\Lambda = 0 \nonumber
\end{eqnarray}
where $x=\Omega(z)-1$.  These are accurate to 1\% in the range
$\Omega(z) = 0.1$-$1$.

If the distribution of the baryonic gas is also isothermal we can
define a ratio of the `temperature' of the collisionless material
($T_\sigma = \mu m_p \sigma^2 / k)$ to the gas temperature:
\begin{equation}
\beta = \frac{\mu m_p \sigma^2}{ k T}.
\label{eq:chapter6_beta}
\end{equation}
We take $\mu = 0.59$.  Given equations~(\ref{eq:mass_sigma_a}) and
(\ref{eq:chapter6_beta}), the relation between temperature and mass is:
\begin{eqnarray}
kT & = & \frac{G M^{2/3} \mu m_p}{2 \beta} 
	\left(\frac{H^2(z) \Delta_c}{2G}\right)^{1/3} \\
   & = & 1.39 f_T
	\left(\frac{M}{10^{15} M_\odot}\right)^{2/3}
 	\left(h^2 \Delta_c E^2 \right)^{1/3} \hbox{keV} 
\label{eq:temp_mass}
\end{eqnarray}
Since we will test the relations~\ref{eq:mass_sigma},
\ref{eq:chapter6_beta} and \ref{eq:temp_mass} against the numerical
simulations separately, we also add a normalization factor,
$f_T$, and set $\beta=1$ for this last equation.

We can easily find the scaling behaviour of a cluster's X-ray
luminosity by assuming bolometric Brems\-strah\-lung emission and
ignoring the temperature dependence of the Gaunt factor (e.g.
\cite{spi78}): $L_{bol} \propto M \rho T^{1/2}$.

We could compute the luminosity by 
using the isothermal sphere approximation, however this is either
infinite, if there is no rollover in density as $r \rightarrow 0$, or we
must arbitrarily select a core radius.  Instead, we will assume the
normalization found by NFW (adjusted slightly to match our definition
of the virial mass):
\begin{equation}
L_{bol} = 1.3 \times 10^{45} 
	\left(\frac{M}{10^{15} M_\odot}\right)^{4/3} \\
	\left(h^2 \Delta_c E^2 \right)^{7/6}
	\left(\frac{\Omega_b}{\Omega}\right)^2 \hbox{erg/s},
\label{eq:mass_lumin_nfw}
\end{equation}
To obtain this we have used the redshift dependence of the critical
density, the temperature-mass relation and multiplied by the 
mean baryon fraction.

Other scaling laws can easily be derived from these; we will need one
more:
\begin{equation}
L_{bol} = 6.8\times 10^{44} 
        \left(\frac{kT/f_T}{1.0\hbox{ keV}}\right)^2
	\left(h^2 \Delta_c E^2 \right)^{1/2}
	\left(\frac{\Omega_b}{\Omega}\right)^2 \hbox{erg/s}.
\label{eq:temp_lumin_nfw}
\end{equation}
We note that cosmology enters these relations only with the
combination of parameters $h^2 \Delta_c E(z)^2$, which comes from the
relation between the cluster's mass and mean density.  The redshift
variation comes mostly from $E(z)$ which is equal to $(1+z)^{3/2}$ for
an Einstein-de Sitter universe.


\subsection{Resolution effects}
\label{sec:resolution_effects}

While we do not expect the first three bulk properties (mass,
temperature and velocity dispersion) to be strongly dependent on the
numerical resolution, the X-ray emissivity is very sensitive to the
density profile of the cluster since it originates primarily from the
central region of the cluster.  In order to examine
the expected luminosity behaviour with resolution, we take the density
profile fit from high-resolution N-body simulations found in
Navarro, Frenk \& White (1996)\markcite{nav96}: 
\begin{equation}
\frac{\rho(r)}{\rho_{crit}} = \frac{\delta_0/c^3}{x(1/c + x)^2},
\label{eq:profile_nfw}
\end{equation}
where $x = r/r_{vir}$.  For clusters, a reasonable parameter choice is
$c = 5$, although our results will not be sensitive to
small changes in this value.  This is a fit to the collisionless
component and not the baryonic gas (as would be more appropriate),
however the difference only becomes important at $r/r_{vir} \leq
0.04$ (c.f. \cite{NFW}; \cite{bry97a}) which will not substantially
affect our results.

In order to approximate the effect of finite numerical resolution we
filter the density distribution:
\begin{equation}
\rho^\prime(r, r_{sm}) = \frac{\int \rho(\vec{r}^\prime)
         W_G\left(\frac{|\vec{r}^\prime-\vec{r}|}{r_{sm}}\right) d^3 
		\vec{r}^\prime}
   {\int W_G\left(\frac{|\vec{r}^\prime-\vec{r}|}{r_{sm}}\right) d^3 
		\vec{r}^\prime}.
\label{eq:dens_smooth}
\end{equation}
The smoothing kernel is a Gaussian ($W_G(x) = e^{-x^2/2}$) and
$r_{sm}$ is the smoothing radius.
Using symmetry, this integral can be partially computed analytically:
\begin{equation}
\rho^\prime
 = \frac{1}{\sqrt{2\pi} r r_{sm}}
	\int_0^\infty \rho(r^\prime) r^\prime 
	\left( e^\frac{-(r^\prime-r)^2}{2 r_{sm}^2} -
	       e^\frac{-(r^\prime+r)^2}{2 r_{sm}^2} \right) dr^\prime.
\end{equation}

\begin{figure}
\epsfxsize=3in 
\centerline{\epsfbox{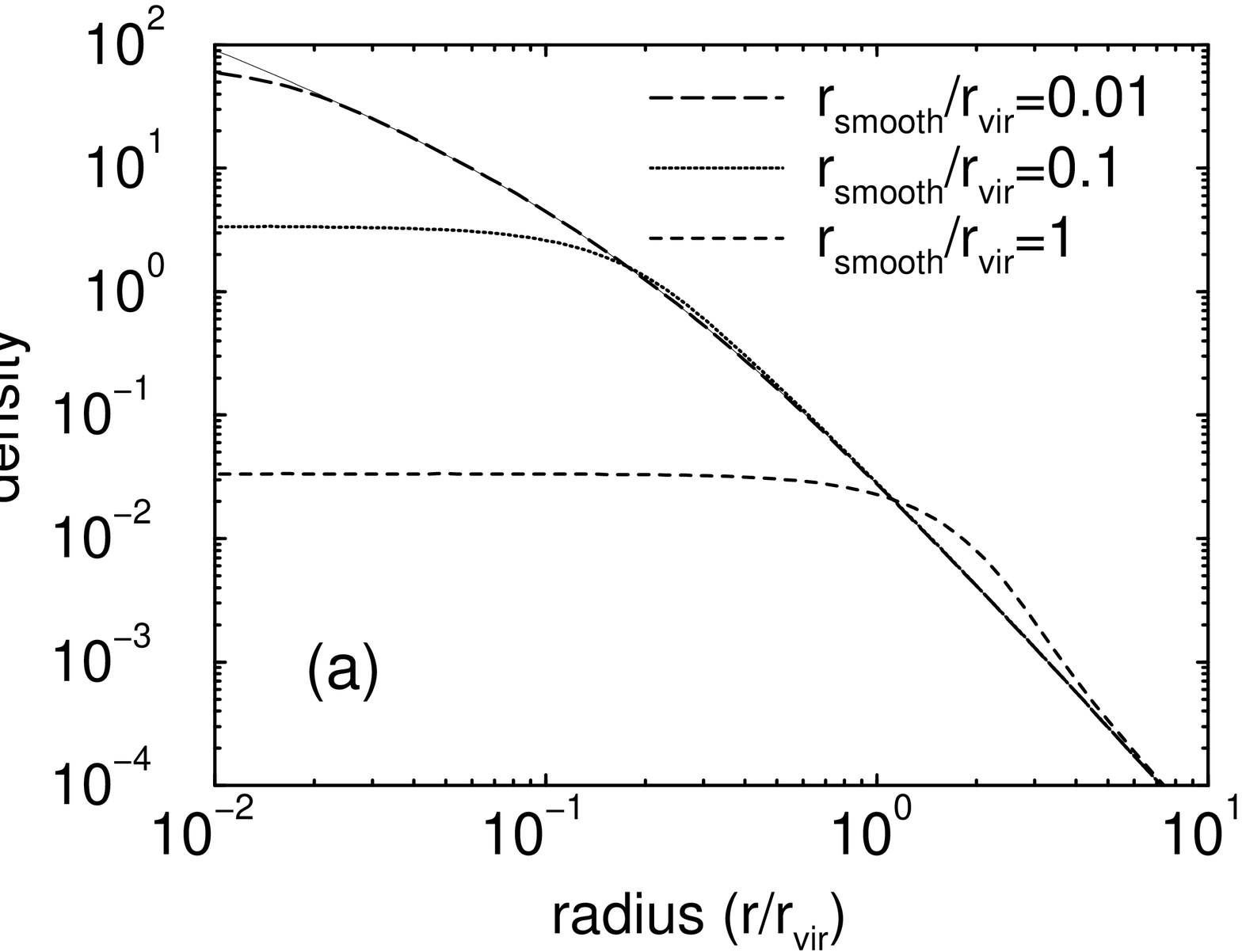}}
\epsfxsize=3in 
\vspace{0.2in}
\centerline{\epsfbox{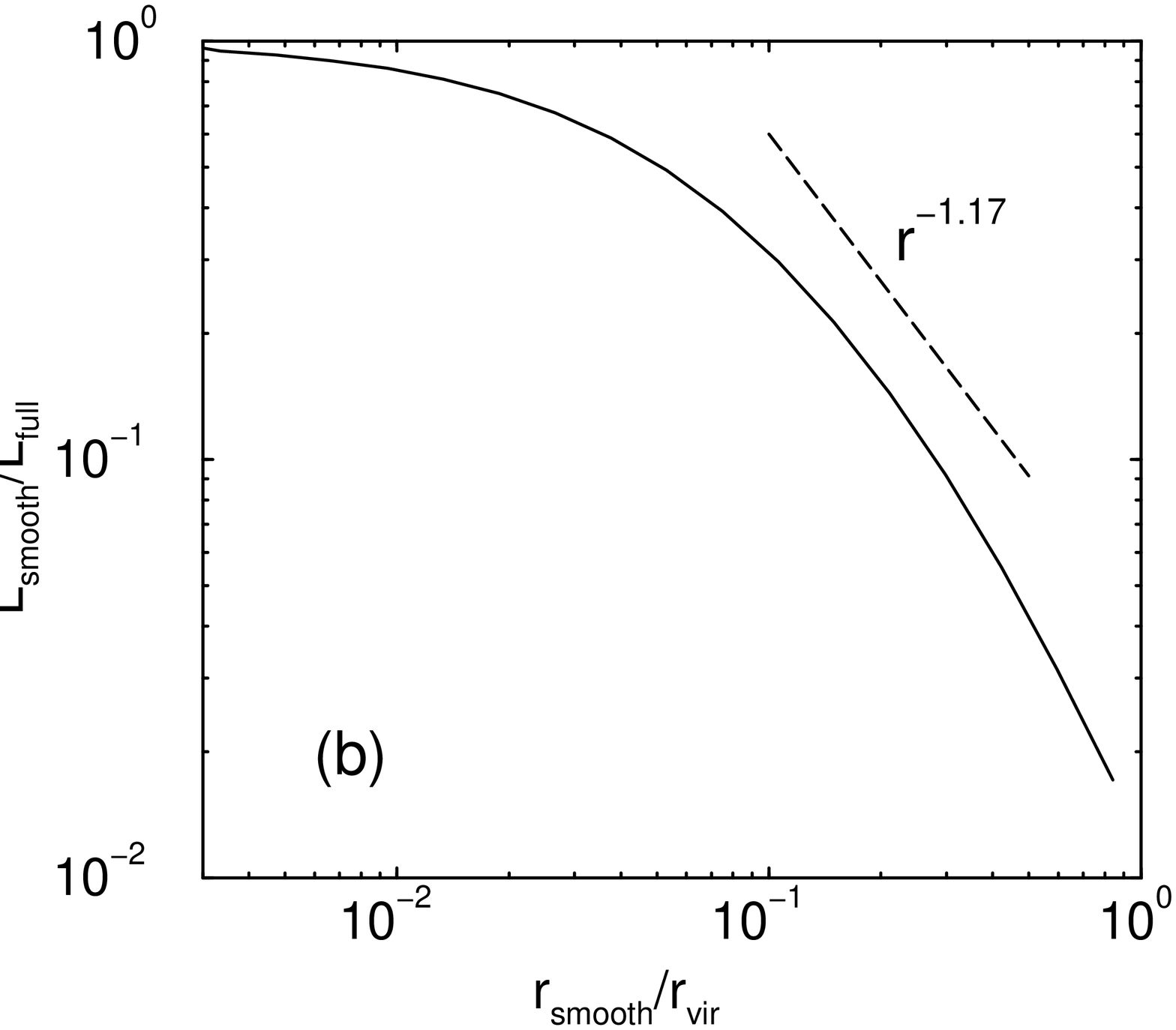}}
\caption{
(a) The density profile in equation~(\protect\ref{eq:profile_nfw})
smoothed by a Gaussian filter with three different values of the
smoothing radius.
(b) The relative bolometric bremsstrahlung luminosity of the smoothed
density profile as a function of smoothing radius.  The dashed line
shows the result found by direct numerical simulation 
(AN96).
}
\label{fig:lumin_smooth}
\end{figure}

In Figure~\ref{fig:lumin_smooth}a, we show the effects of various
smoothing radii to the adopted density profile.  The mass
is conserved by the smoothing process in the sense that the total
mass within some fixed fraction $r/r_{sm} \gg 1$ is constant as
$r_{sm}$ changes (the total mass of equation~(\ref{eq:profile_nfw})
does not converge as $r \rightarrow \infty$).  This results in mass
being transferred from small to large radii, causing the profile to
steepen outside the smoothing radius, another feature observed by
AN96.  Assuming the temperature profile does not change, the
bolometric free-free luminosity can be computed as a function of
$r_{sm}/r_{vir}$ and the result is shown in Figure~\ref{fig:lumin_smooth}b.
Also shown is the scaling result found by AN96 and given by
equation~(\ref{eq:lumin_resolution}), which they found to be
approximately valid in the range $0.1 > r_{sm}/r_{vir} > 1$.  The
agreement is good enough that we will adopt their value ($\alpha =
1.17$), although it should be kept in mind that this power-law
behaviour is only approximately correct.


If all density profiles can be scaled by the virial radius and critical
density to agree with equation~(\ref{eq:profile_nfw}) regardless of mass
(at least in a statistical sense), then the smoothed luminosity will
scale in general as $L_{bol} \propto (r_{vir}/r_{sm})^{\alpha}$.
Recent work (\cite{nav96}; \cite{col96}) indicates that, in fact, this
is not true and lower mass objects form earlier and have steeper profiles.
Nevertheless it is a good approximation over the range of
masses examined here and we will adopt it. 

Using the relation between the virial radius and virial mass, 
we can include the effects of numerical
resolution in equation~(\ref{eq:mass_lumin_nfw}):
\begin{eqnarray}
L^\prime_{bol} & = & 5.8 \times 10^{45} 
	\left(\frac{M}{10^{15} M_\odot}\right)^\frac{4+\alpha}{3}
	\left(h^2 \Delta_c E(z)^2 \right)^\frac{7-2\alpha}{6} 
	\nonumber \\ & &
	\left(\frac{\Omega_b}{\Omega}\right)^2 
	\left(\frac{100 \hbox{ kpc}}{\Delta x}\right)^\alpha
	\left(1+z\right)^{\alpha} \hbox{ erg/s},
\label{eq:mass_lumin_code}
\end{eqnarray}
Here we have written the resolution in terms of the comoving cell size
$\Delta x$ and normalized by assuming that $\Delta x = 100$ kpc will
correctly reproduce the observed core radius of a $10^{15} M_\odot$
cluster (c.f. AN96).  The resulting $L^\prime_{bol}$-$T$
relation is,
\begin{eqnarray}
\label{eq:temp_lumin_code}
L^\prime_{bol} & = & 2.5\times 10^{45} 
        \left(\frac{kT/f_T}{1.0\hbox{ keV}}\right)^\frac{4+\alpha}{2}
	\left(h^2 \Delta_c E(z)^2 \right)^\frac{1-\alpha}{2}
	\nonumber \\ & &
	\left(\frac{\Omega_b}{\Omega}\right)^2 
	\left(\frac{100 \hbox{ kpc}}{\Delta x}\right)^\alpha
	\left(1+z\right)^{\alpha} \hbox{ erg/s}.
\end{eqnarray}


\subsection{Comparison to simulations}

Our numerical techniques have been described in detail elsewhere
(\cite{PPM}), so we only briefly summerize them here.  The dark
component was modelled by the particle-mesh method with the Poisson
equation solved through an FFT.  The adiabatic equations of gas
dynamics are solved with a modified version of the piecewise parabolic
method.  Both techniques have a resolution of a few cell widths.
Radiative cooling is not included as we have insufficient resolution
to properly follow the cooling structures; however, for large clusters,
which have very long cooling times over most of their volume, this may
be a reasonable approximation.

The four simulations we use in this paper are summarized in
Table~\ref{table:simulations}.  The results from the CHDM256 model are
sufficiently similar to CHDM512 that we mostly focus on the first
three models.  The resolution, as measured by cell width, is similar
for the three 85$h^{-1}$ Mpc boxes, while substantially better for
the CHDM512 model, giving us some leverage in studying the effects of
resolution.
All have power spectrum normalization (indicated here by the
$\sigma_8$, the size of mass fluctuations in spheres of radius
8$h^{-1}$ Mpc) which are approximately in agreement the the amplitude
of fluctuations in the cosmic background radiation on large scales.
Some of these simulations and their initializations have been
described elsewhere (\cite{cdm_xray}; \cite{chdm_xray}).


{
\begin{deluxetable}{cccccccccc}
\footnotesize
\tablecaption{Simulation parameters}
\tablewidth{0pt}
\tablehead{
\colhead{designation} & \colhead{$\Omega_{\hbox{cold}}$} &
\colhead{$\Omega_{\hbox{hot}}$} & \colhead{$\Omega_{\hbox{baryon}}$} & 
\colhead{h} & \colhead{$m_{\nu}$ (eV)} & \colhead{$\sigma_8$} & 
\colhead{$N_{\hbox{cell}}$} & \colhead{$N_{\hbox{part}}$} &
\colhead{$L_{\hbox{box}}$ ($h^{-1}$Mpc)}
}
\startdata
CDM270  & 0.94  & 0.0 & 0.06  & 0.5 & 0              & 1.05 & $270^3$
& $135^3$ & 85 \nl
CHDM512 & 0.725 & 0.2 & 0.075 & 0.5 & $2 \times 2.3$ & 0.7  & $512^3$ 
& $3 \times 256^3$ & 50 \nl
OCDM270 & 0.34  & 0.0 & 0.06 & 0.65 & 0              & 0.75 & $256^3$
& $128^3$       & 85 \nl
CHDM256 & 0.6   & 0.3 & 0.1   & 0.5 & 7.0            & 0.67 & $256^3$ 
& $3 \times 128^3$ & 85 \nl  
\enddata
\label{table:simulations}
\end{deluxetable}
}


To identify clusters we adopt the spherical overdensity algorithm
described in Lacey \& Cole (1996)\markcite{lac96}.  Peaks in the
density distribution are identified as cluster centers; spheres are
grown around each point until the mean interior density reaches
$\Delta_c$.  The center-of-mass inside the region is computed and this
points is used to grow a new sphere.  The procedure is iterated to
convergence and then all clusters are checked for overlap (defined as
a center-to-center distance less than three-quarters of the sum of
their virial radii), with the less massive cluster being removed.  The
scaling relations are robust to changes in what we do with overlapping
clusters.

In the following sections we examine each scaling law in detail and
compare to other work where possible.


\subsubsection{The $M$-$\sigma$ relation}

\begin{figure*}
\epsfxsize=5.5in
\centerline{\epsfbox{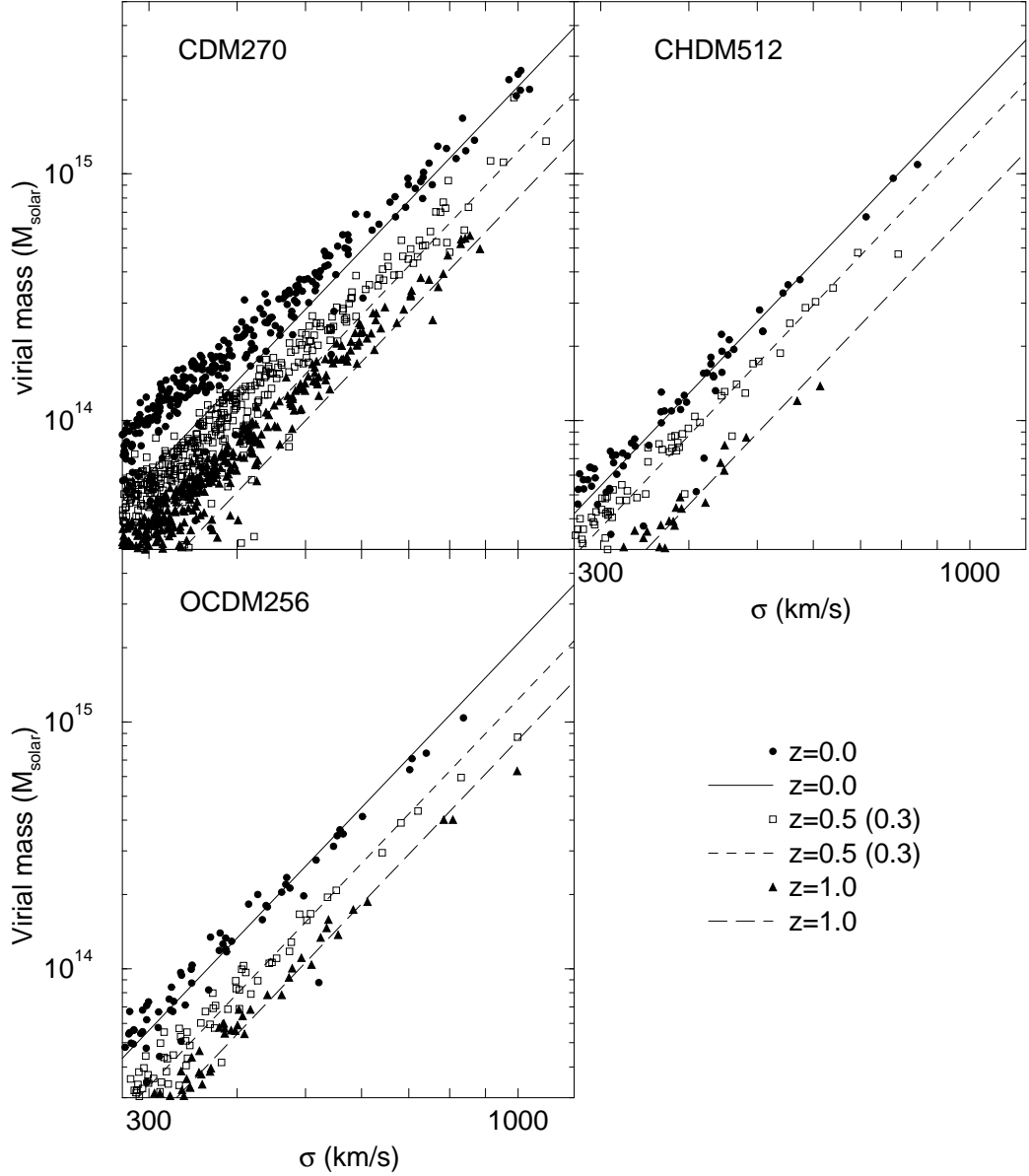}}
\caption{
The virial mass and one-dimensional velocity dispersion of clusters
identified in the three models for three different redshifts:
$z=0$, $0.5$ ($0.3$ for CHDM512) and 1 along with the scaling relation
from equation~(\protect\ref{eq:mass_sigma}) with $f_\sigma$ given by
Table~\protect\ref{table:scaling_fits}.
}
\label{fig:mass_sigma}
\end{figure*}

In Figure~\ref{fig:mass_sigma}, the mass-weighted one-dimensional
velocity dispersions and virial masses are plotted for clusters
identified in our three primary models, at three different redshifts:
$z=0$, $0.5$ and 1 (except for the CHDM512 simulation, for which the
$z=0.5$ output was corrupted so $z=0.3$ was used instead).  The
velocity dispersion includes both cold and hot collisionless
components, where appropriate, and is computed after removing the
center-of-mass velocity.  There are substantially more clusters in the
CDM270 simulation because of its higher normalization (the CHDM512
run also used a somewhat smaller volume).

We also show the predicted virial relation between
these two quantities, given by equation~(\ref{eq:mass_sigma}).  The
normalization, specified by $f_\sigma$, is determined separately for each
model by fitting the 25 most massive clusters at $z=0$; the numerical
values are listed in Table~\ref{table:scaling_fits}, along with their
standard deviations.  The results are almost completely insensitive to
small (factor of two) changes in the number of clusters used.  The
same normalization is used at higher redshifts.  For the most massive
--- and best resolved --- clusters, the agreement in both slope and
redshift variation is quite good, with remarkably little scatter.  For
less well-resolved clusters, those with $\sigma$ below 600 km/s, the
slope flattens.  This is most likely due to resolution effects, as the
virial radius drops to just a few cell widths.

\begin{deluxetable}{cccccccccc}
\footnotesize
\tablecaption{Virial fitting parameters}
\tablewidth{0pt}
\tablehead{
\colhead{source} & \colhead{$f_\sigma$} & \colhead{$\Delta \sigma$} &
\colhead{$f_T$} & \colhead{$\Delta T$} & \colhead{$N_{\hbox{clust}}$}
}
\startdata
CDM270  & 0.85 & 0.05 & 0.79 & 0.10 & 25 \nl
CHDM512 & 0.89 & 0.03 & 0.75 & 0.14 & 25 \nl
OCDM256 & 0.85 & 0.04 & 0.78 & 0.05 & 25 \nl
CHDM256 & 0.82 & 0.04 & 0.75 & 0.05 & 25 \nl
E91     & 0.98 & 0.02 & 0.81 & 0.05 & 22 \nl
NFW     & 0.96 & 0.04 & 0.86 & 0.11 & 6 \nl
CG95    & 0.91 &      &      &      & $\sim$50 \nl
CL96    &$\sim1$&     &      &      & $\sim$100 \nl
EMN96   &      &      & 0.92 &$\sim$0.05& 58 \nl
\enddata
\label{table:scaling_fits}
\end{deluxetable}

\begin{figure}
\epsfysize=2.9in
\centerline{\epsfbox{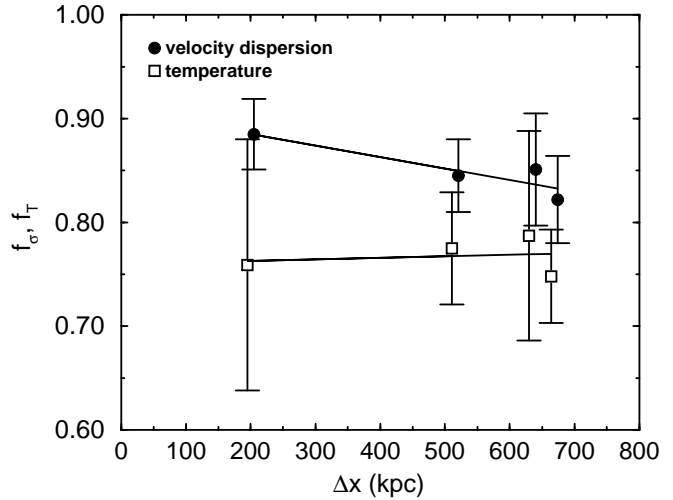}}
\caption{
Normalizations for the M-T and M-$\sigma$ virial relations (as
determined by numerical simulation) as a function of resolution,
measured by cell size in kpc at $z=0$.  The $f_\sigma$ points have
been slightly offset in $\Delta x$ for clarity.
Lines are the best-fit linear relation.
}
\label{fig:fit_resolution}
\end{figure}

To examine any systematic variation with resolution, the normalization
from each model is plotted against the cell size of that simulation
in Figure~\ref{fig:fit_resolution} (here we include the CHDM256
model).  Shown as an error bar is the scatter, parameterized by
each sample's standard deviation.  Although not very significant, the
best fitting straight line shows a modest increase in $f_\sigma$
with increasing resolution.  This is in the expected direction since
improved force resolution requires higher velocity dispersions to
maintain equilibrium.

In Table~\ref{table:scaling_fits}, we also list the normalizations
determined by other authors, using different numerical techniques.
Evrard (\cite{evr91}, E91) performed 22 simulations of isolated
clusters in an Einstein-de Sitter universe with a cold dark matter
spectrum and obtained a somewhat larger normalization.  The six
($\Omega = 1$) clusters of NFW, using a TREESPH code, also had a
slightly larger normalization.  They used a marginally different
virial overdensity, however when fitting their results to
equation~(\ref{eq:mass_sigma}), we adopt their value of $\Delta_c =
200$.  The effect of increasing this parameter is to sample a slightly
smaller volume, however since the velocity-dispersion profile is
nearly flat in this range (c.f. section~\ref{sec:profiles}), the
effect is very small.  The P$^3$M simulations of Crone and Geller
(1995, CG95)\markcite{cro95} resulted in a mean $f_\sigma = 0.91$ (after
converting from their notation) for well resolved clusters, with
$\Delta_c = 300$.  This also found that the smallest clusters deviated
slightly from the virial power-law, ascribing the effect, as we do, to
insufficient numerical resolution.  Their scatter appears similar to
that found here.  Finally, the P$^3$M simulations of Cole \& Lacey
(1996, CL96) produced similar results.  They did not fit the virial
relation, but their Figure~3 indicates both that the virial relation
is a reasonable fit and that there is not substantial variation from a
power law over the range of masses examined.  Jing \& Fang
(1994)\markcite{jin94} argued for $f_\sigma = 0.8$--0.9, but this was
based on combined N-body and Press-Schechter comparisons, rather than
directly from the $M-\sigma$ relation.  Finally, a number of other
authors have studied this issue but often in ways which make it
difficult to compare directly to the virial formalism.  For example,
Walter \& Klypin (1996) used a fixed cluster radius rather than one
based on the mean overdensity.

In section~\ref{sec:resolution_effects}, we mentioned that
high-resolution N-body simulations indicate that the dark matter
density profile does not follow the isothermal $r^{-2}$ profile but is
instead given by equation~(\ref{eq:profile_nfw}), which has two
parameters.  If one ($\delta_0$) is set by the requirement that the
mean density inside the virial radius is $\Delta_c$, the other is
strongly correlated with mass (\cite{nav96}), so the mean radial
profile is completely specified by the mass (or, more accurately, the
redshift of formation).  The implied value of $f_\sigma$ (i.e. the
correction as compared to the isothermal profile), can be computed
using Jeans equation ($\rho^{-1} d[\rho\sigma(r)^2]/dr = - d\phi/dr$;
we assume the velocity dispersion is isotropic for simplicity).  It is
the ratio of the mass weighted velocity dispersion (squared) to the
isothermal virial value:
\begin{eqnarray}
f_\sigma^{2} &=& \frac{\bar{\sigma}^2}{\sigma_{vir}^2} =
	 \frac{2c}{\left(\ln\left(1+c\right) - \frac{c}{1+c}\right)^2}
	 \int_0^c y^2 dy
	\nonumber \\ & & \hspace*{-0.3cm}
	 \int_y^\infty \frac{dx}{x^3\left(1+x\right)^2}
	 \left[\frac{1}{1+x} - 1 + \ln\left(1+x\right)\right]
\end{eqnarray}
For the range of masses and cosmologies studied here, c varies from
5-10, and the resulting $f_\sigma$ = 0.92-0.97.  

Thus we have strong evidence from a variety of sources that the virial
relation describes clusters fairly well, although the normalization
may need to be slightly modified, with a value around $f_\sigma \sim
0.9$-0.95 in good agreement with a wide variety of numerical methods.
Unfortunately, there is likely to be a systematic bias between the
velocity dispersion of the dark matter and that of the galaxies
(\cite{car90}; \cite{sum95}) which we cannot address with these
simulations, so we turn to the thermal temperature of the gas for a
better observational diagnostic.


\subsubsection{The $M$-$T$ relation}

\begin{figure*}
\epsfxsize=6in
\centerline{\epsfbox{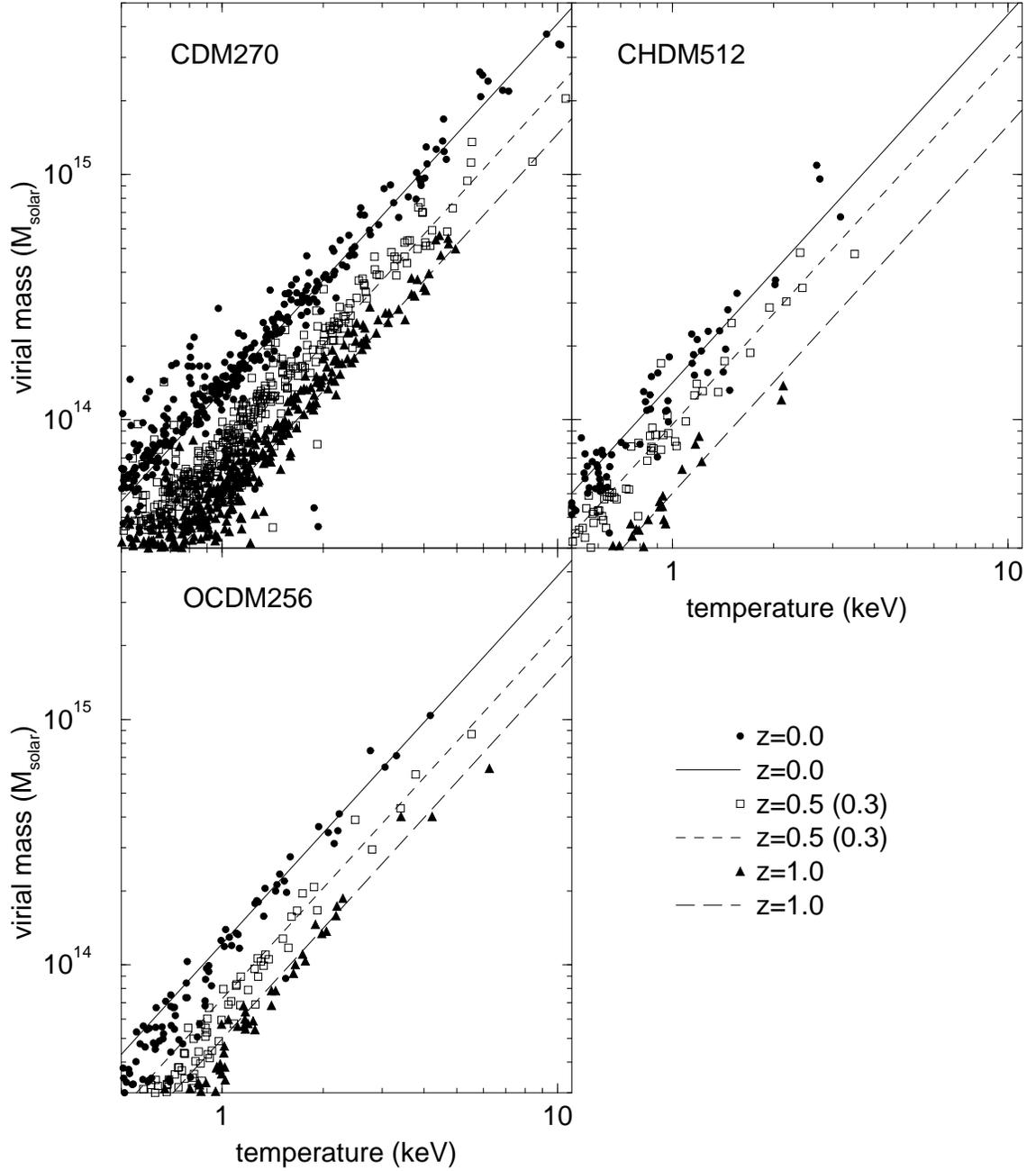}}
\caption{
The virial mass and emissivity-weighted temperature of clusters
for the same models and redshifts as in
Figure~\protect\ref{fig:mass_sigma}.  Lines are the virial scaling
relations from equation~(\protect\ref{eq:temp_mass}), with
normalizations given in Table~\protect\ref{table:scaling_fits}.
}
\label{fig:mass_temp}
\end{figure*}

The temperature-mass relation for our models is shown in
Figure~\ref{fig:mass_temp}.  The temperature is a bolometric-
luminosity weighted average across the cluster.
In contrast to the $M$-$\sigma$ relation, even the most poorly resolved
clusters follow the virial relation, with the
exception of the open model, in which very low mass clusters have a
higher temperature than predicted.  This may be due to the earlier
formation times for these clusters, as compared to the other models
(\cite{nav96}).

As before, the virial relations are fit with the 25 most massive
clusters at $z=0$, and the results, as parameterized by $f_T$ are
given in Table~\ref{table:scaling_fits} and plotted against cell size
in Figure~\ref{fig:fit_resolution}.  There is no discernible trend with
resolution.

In the same Table, we also show the value of $f_T$ as determined by
other simulations, all of which used smoothed particle hydrodynamics
(SPH).  The cluster simulations of Evrard (1991) produced a
normalization very close to that obtained here, although he used a
mass-weighted temperature.
The six clusters of NFW again indicate a slightly higher
value, although still within 10\%. 
Recently, Evrard, Metzler \& Navarro (1996, EMN96) examined
simulations of 58 clusters (a sample which included the six of NFW)
for a variety of models, including those with low $\Omega$.  For
$\Delta_c = 250$, they find a mean value of $f_T = 0.92$.  This is
17\% above our mean result.  A careful examination of their Figure~4
indicates that there are slight systematic differences between the
simulations, with clusters in flat universes implying somewhat lower
values for $f_T$, and $\Omega=0.2$ (flat and open) models indicating
higher normalizations.  They also included a set of clusters which
included substantial galactic winds, which appears to increase the
virial normalization only very slightly.  In it is interesting to note
that the other physical contribution which we have not included here,
cooling flows, would tend to decrease $f_T$, although the expected
effect is quite small.

The range exhibited in
Table~\ref{table:scaling_fits} is surprisingly low, considering the
range of models, resolutions, numerical techniques and researchers who
have examined this robust relation.  The size of the scatter is also
quite low, around 10\% (corresponding to a 15\% scatter in mass).
This implies that the temperature is an excellent virial mass
indicator, even better than using $\beta$-model estimates
(\cite{evr96}).


\subsubsection{The $T$-$\sigma$ relation}

\begin{figure*}
\epsfxsize=6in
\centerline{\epsfbox{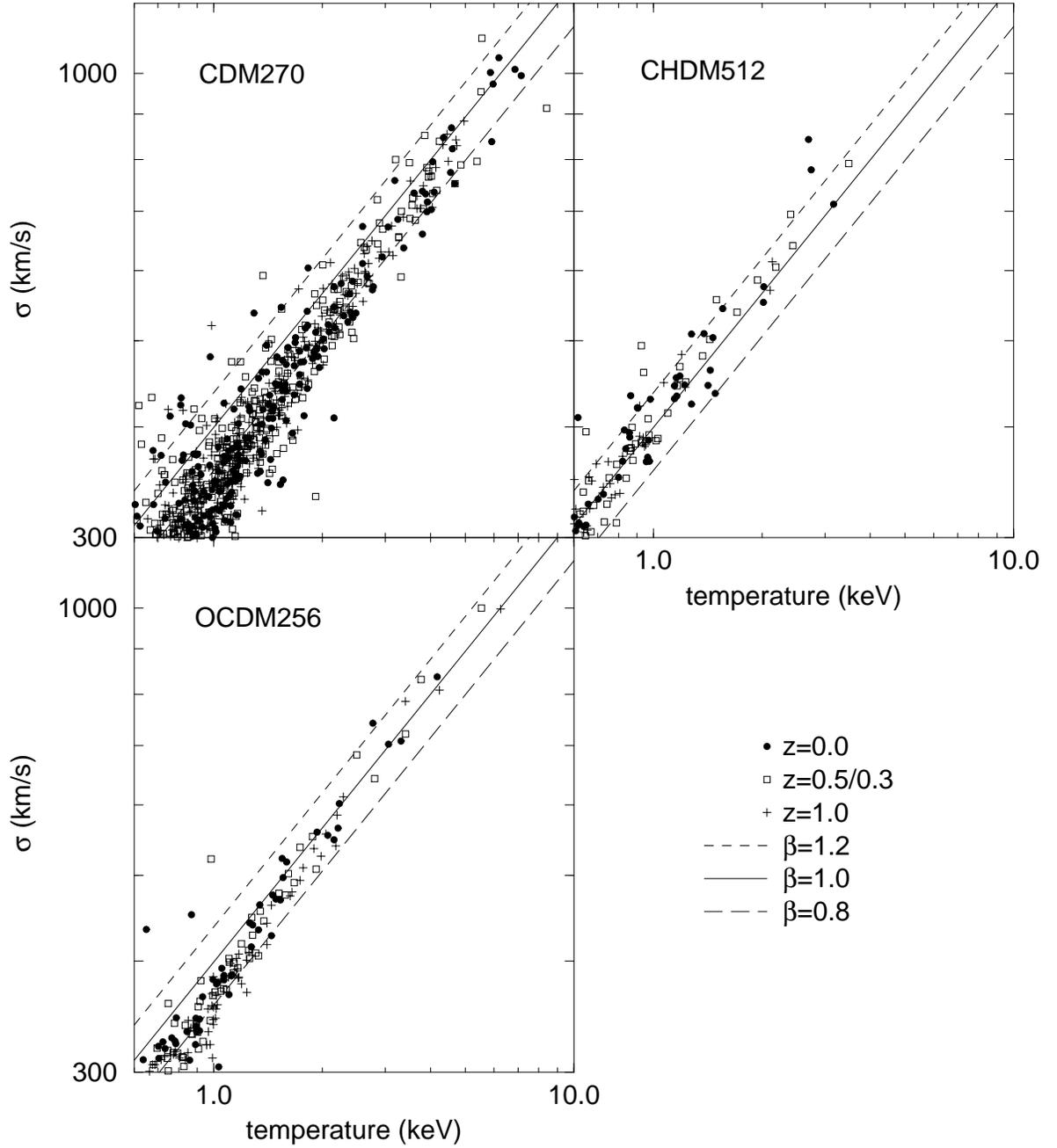}}
\caption{
The velocity-dispersion and emissivity-weighted temperature of
clusters identified for the same models and redshifts as in
Figure~\protect\ref{fig:mass_sigma},
along with the scaling relation from
equation~(\protect\ref{eq:chapter6_beta}).
}
\label{fig:temp_sigma}
\end{figure*}

The relation between temperature and velocity dispersion is plotted in
Figure~\ref{fig:temp_sigma}, along with
equation~(\ref{eq:chapter6_beta}) for three values of $\beta$
(0.8,1,1.2).  All simulations show curves which are steeper than
predicted and there is a trend with mass, moving from low to high to
values of $\beta$.  This is probably a reflection of the fact that low
mass clusters are more poorly resolved and so have a lower $\sigma$
than predicted (see also the $M$-$\sigma$ relation).  The best
resolved clusters, in fact, have values above one.  This follows from
the $M$-$T$ and $M$-$\sigma$ relations, since $\beta = f_\sigma^2/f_T
\approx 1.06$.  This value is close to the mean of NFW (1.07) but
somewhat lower than the 1.18 found by Evrard (1991).  Our result
should most likely be viewed as a lower limit, since
Figure~\ref{fig:fit_resolution} indicates that the $\sigma$ values
have not yet converged.
The most likely
explanation for $\beta > 1$ is incomplete thermalization in the gas,
although since the two value are computed with different weights, this
is not completely straightforward.

Observationally, the quantity $\beta$ has been the subject of
some investigation, with values being generally compatible with one
(\cite{gir96}; \cite{lub93}), although this is for the galaxy velocity
dispersion rather than the collisionless component.  See also Lubin \etal 
(1996)\markcite{lor96}.


\subsubsection{The $M$-$L_{bol}$ relation}

\begin{figure*}
\epsfxsize=6in
\centerline{\epsfbox{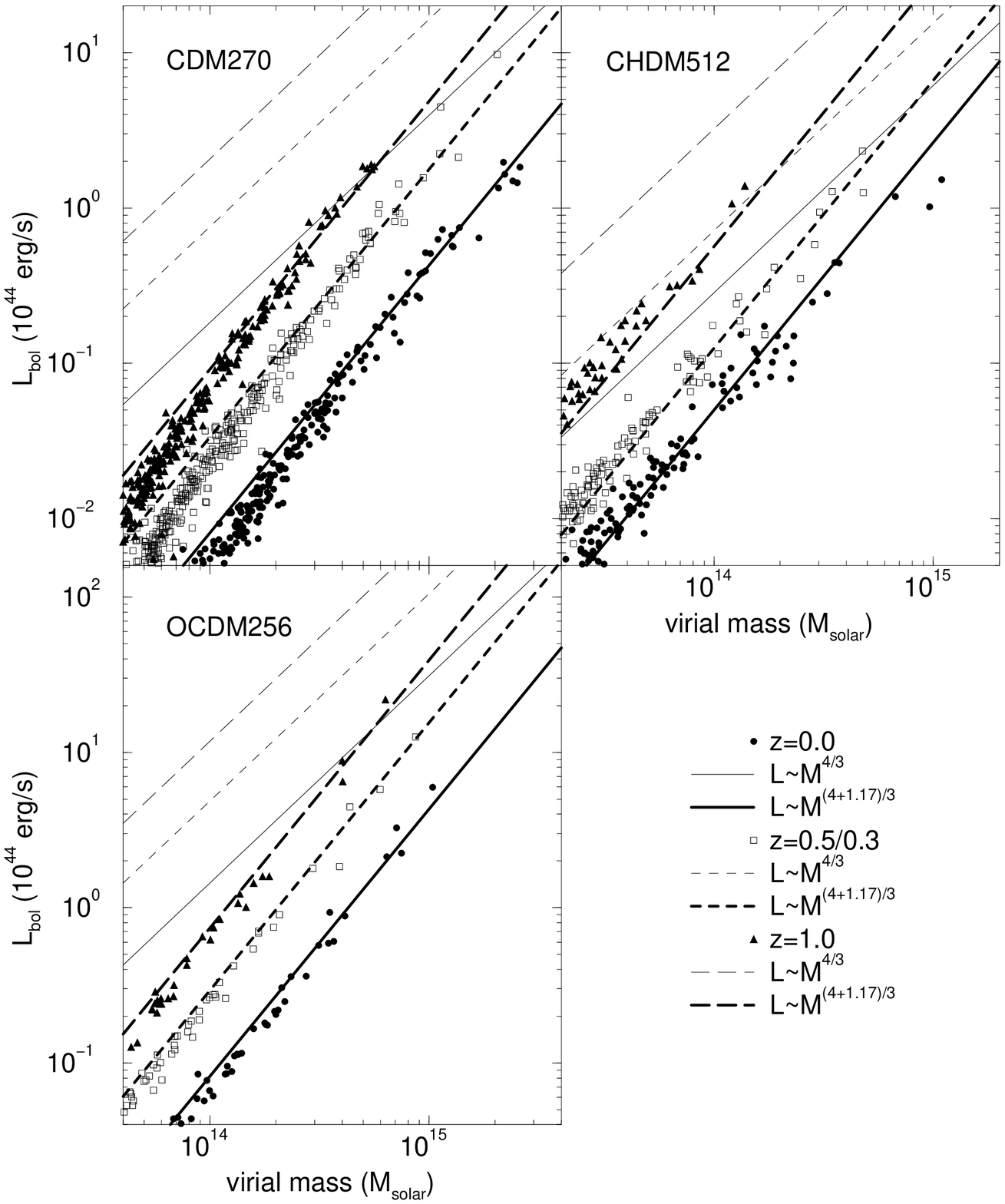}}
\caption{
The mass and bolometric luminosity of
clusters identified for the same models and redshifts as in
Figure~\protect\ref{fig:mass_sigma}.  Thin curves are the virial
scaling relation from equation~(\protect\ref{eq:mass_lumin_nfw}),
while thick lines are from the resolution-adjusted version, 
equation~(\protect\ref{eq:mass_lumin_code}). 
}
\label{fig:mass_lumin}
\end{figure*}

Figure~\ref{fig:mass_lumin} shows the mass-bolometric luminosity
relation for our three models.  Shown are both the virial predictions
from equation~(\ref{eq:mass_lumin_nfw}) as thin lines, and the
resolution-adjusted equation~(\ref{eq:mass_lumin_code}), as thick
curves.  The latter relation fits the simulated clusters quite well.
As we demonstrated in section~\ref{sec:resolution_effects},
a cluster's luminosity is diminished by a factor which depends on its
mass (and the resolution of the simulation).  Another way to say this
is that the core radius of the cluster is set by the the cell size,
rather than scaling with cluster mass and density.  Although the
slopes differ, the variation in redshift for the two relations is very
similar.  Also, we should note that the curves are not fit (as for the
$M$-$T$ and $M$-$\sigma$ relations), but all have the same, somewhat
arbitrary, normalization given in section~\ref{sec:scaling_review}.

\subsubsection{The $L_{bol}$-$T$ relation}

\begin{figure*}
\epsfxsize=6in
\centerline{\epsfbox{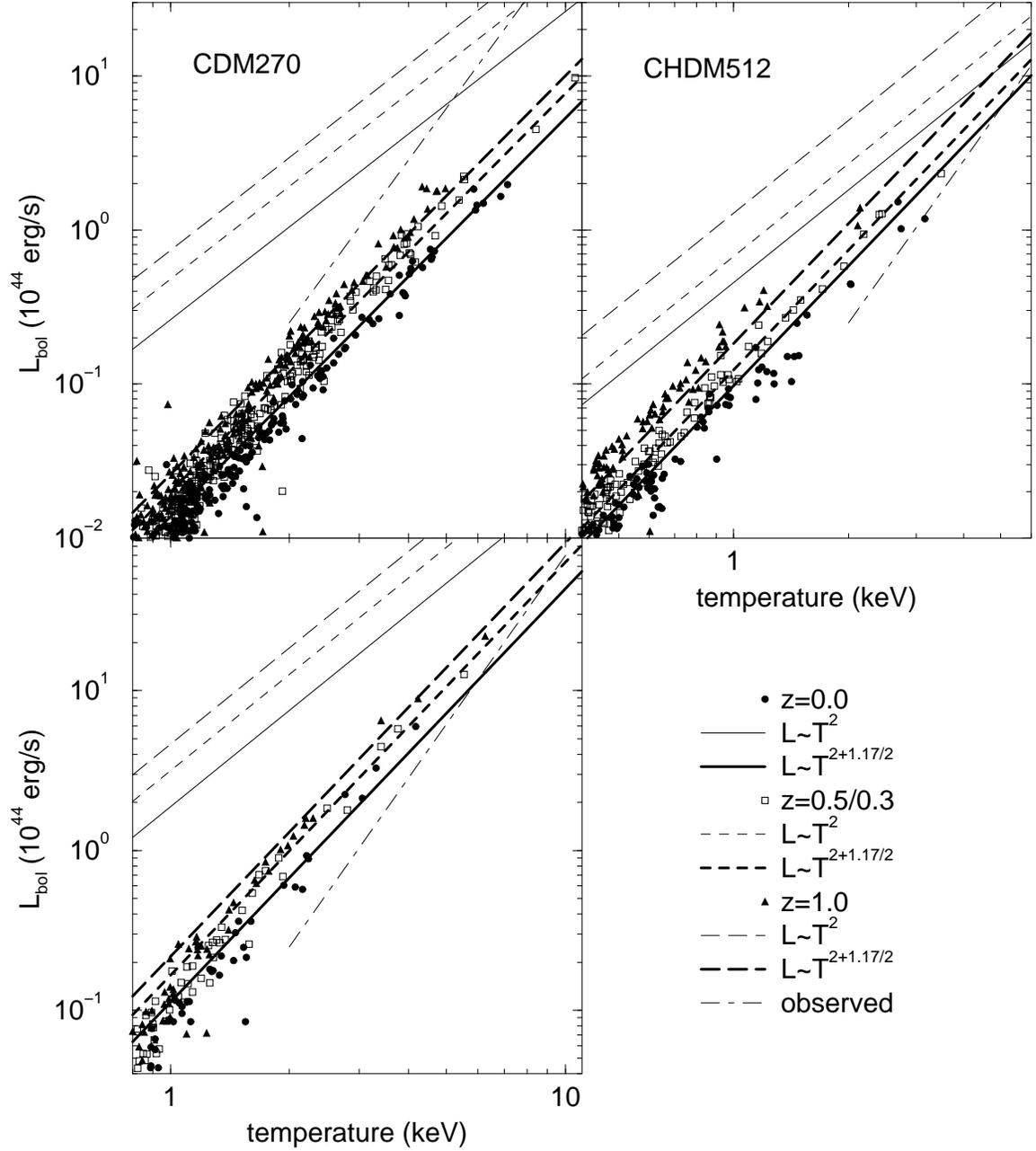}}
\caption{
The temperature and bolometric luminosity of
clusters identified for the same models, redshifts and virial
relations as in Figure~\protect\ref{fig:mass_lumin}.  The dot-dashed
line indicates the observed relation $L \sim T^3$ at $z=0$.
}
\label{fig:temp_lumin}
\end{figure*}

The final comparison is the bolometric luminosity-temperature
relation (Figure~\ref{fig:temp_lumin}).  Although it is not
independent of the previous four, it is the one with the
best observational constraints.  The observations indicate a slope of
$L \sim T^{2.7\hbox{-}3.5}$ (\protect\cite{edg91a},b;
\protect\cite{dav93}).  Although the slope of the
simulated clusters is closer to the observations than that
predicted by the equation~(\ref{eq:temp_lumin_nfw}), we argue that
this is an artifact of our fixed numerical resolution.   This result
applies to Eulerian simulations in general.  Bryan \etal
(1994b)\markcite{chdm_xray} found a slope in agreement with
observations for the 2--10 keV luminosity-temperature (as opposed to the
bolometric result presented here).  As we see here, this was partly
due to resolution effects (but also resulted from the
finite band-pass which strongly curtails the luminosity for clusters
under a few keV).  Thus, the basic discrepancy between the observed
and predicted bolometric $L-T$ relation remains unresolved.
Additional physics, such as galactic feedback, seems to be required
(see also Evrard \& Henry (1991)\markcite{eh91} and NFW).

\subsubsection{Scatter in the virial relations}

The scatter of clusters around the virial relations can be important
because it provides a limit on how accurately masses can be
determined, even in the absence of observational uncertainties.
This is a factor in computing distribution functions of observable
quantities, such as temperature or luminosity, since a given mass is
associated with a range of temperatures.  More concretely, given a
mass distribution $dn/dM$, the resulting temperature function $dn/dT$
(computed by multiplying by the virial relation $dM/dT$) must be
convolved with the scatter.

\begin{figure}
\epsfxsize=3.2in
\centerline{\epsfbox{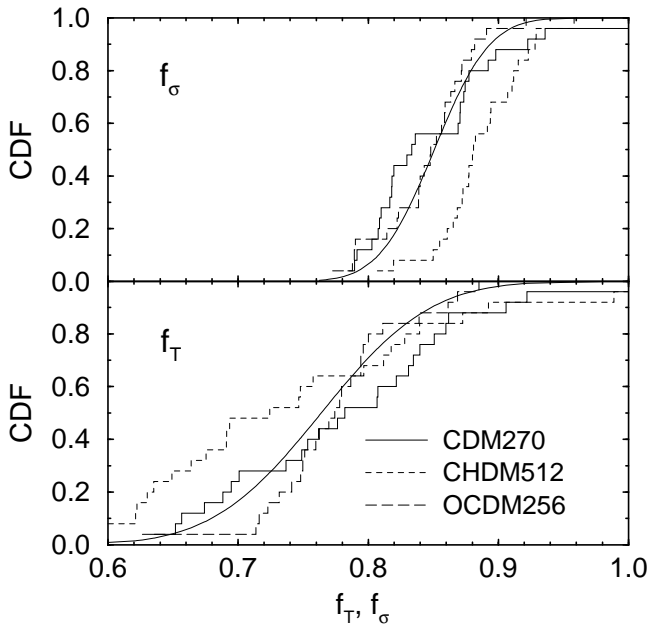}}
\caption{
Cumulative distributions of $f_\sigma$ (top) and $f_T$ (bottom) for
the 25 most massive clusters in our three primary simulations.  The
solid lines are Gaussian distributions fitted to the combined sample.
}
\label{fig:scatter}
\end{figure}

In Figure~\ref{fig:scatter}, we show cumulative distribution functions
for our three canonical simulations.  Although the $f_\sigma$
relation is somewhat tighter than $f_T$, their accuracy in predicting
the cluster mass is nearly identical (since $M \sim \sigma^3 \sim
T^{3/2}$).  As we noted previously, the mean of the velocity dispersion
distribution appears to shift slightly with resolution, but the
temperature does not.  The shape of the distributions are well
described by a Gaussian (as verified by a Kolmogorov-Smirnov test).

In order to compare the size of the fluctuations to those observed in
real clusters, we must use the $L$-$T$ relation.  Although the slope
of our predicted relation disagrees with that observed, we can compare
the scatter around the two curves.  Since real cluster temperatures
also have observational uncertainties, we add the intrinsic scatter
and observed uncertainty in quadrature.  The simulated clusters have a
scatter of $11.4^{+2.3}_{-1.4}$\%.  The flux-limited sample of Henry
\& Arnaud (1991) gives $19.6^{+4.4}_{-2.7}$\% (similar to, but
slightly higher than, the 16.4\% found by Pen, 1996, for the same
sample), while the larger sample of David \etal (1993) requires a
smaller intrinsic scatter: $15.3^{+1.9}_{-1.5}$\%.  These values are
determined by fitting the $L$-$T$ 
relation and adjusting $\sigma_T = fT$ until the resulting $\chi^2$
indicates a good fit; uncertainties are one sigma.  Although the
scatter in the simulated clusters appears to be lower than that
observed, there are two factors which would lead us to overestimate
the observed value.  The first, which is unlikely to be important, is
ignoring the small observational uncertainties in the
luminosity.  More seriously, we have implicitly assumed that the
temperature uncertainties are normally distributed.  Since most
realistic distributions have tail probabilities that are larger than
Gaussian, this causes us to overvalue outliers, forcing the required
intrinsic scatter to rise.  In any case, the difference is not large,
and so we tentatively conclude that the simulations reproduce the
observed scatter. 


\section{Distribution Functions}
\label{sec:distribution}

The scaling relations discussed in the previous section can be
combined with a prescription for computing the mass distribution function of
virialized objects to make predictions of more easily observed
distribution functions.  Although there are many prescriptions
for obtaining the number of collapsed objects given a power spectrum of
linear initial perturbations (e.g. \cite{lac93}; \cite{kit96};
\cite{bon96a}), we adopt the Press-Schechter (PS) formalism
(\cite{pre74}) because it is fairly accurate, relatively simple, and
widely used.

The comoving number of virialized objects of mass $M$ in mass interval
$dM$ is given by,
\begin{equation}
\frac{dn}{dM} = \left(\frac{2}{\pi}\right)^{1/2} \frac{\bar{\rho}}{M} 
	\frac{\nu_c}{\sigma(M)} 
	\frac{d\bar{\sigma}}{dM} \: e^{-\nu_c^2/2}.
\label{eq:press_schechter}
\end{equation}
where $\bar{\rho}$ is the mean density and $\nu_c =
\delta_c/\sigma(M)$.  The
linear rms density fluctuations of the 
power spectrum is given by 
\begin{equation}
\sigma^2(M) = \int W_T^2(kR) P(k) d^3 k
\label{eq:sigma}
\end{equation}
on the scale $M = 4\pi R^3 \bar{\rho}/3$ with a top-hat smoothing
filter $W_T(x) = 3(\sin(x)/x - \cos(x))/x^2$ [note the difference
between the velocity dispersion $\sigma$ and the rms density
fluctuation $\sigma(M)$].  The numerical value of $\delta_c$ has been
the topic of some debate.  The spherical collapse model gives
$\delta_c$ as a slowing varying function of cosmology which is well fit
by: $\delta_c = 1.686\Omega(z)^c$, where $c=0.055$ (0.018) for flat
(zero cosmological constant) models.  However, given the shortcomings
of the theory's theoretical underpinnings (\cite{bon91}), it is
probably a better approach to take $\delta_c$ as a free parameter and
fit for it.

\begin{figure*}
\epsfxsize=4in
\centerline{\epsfbox{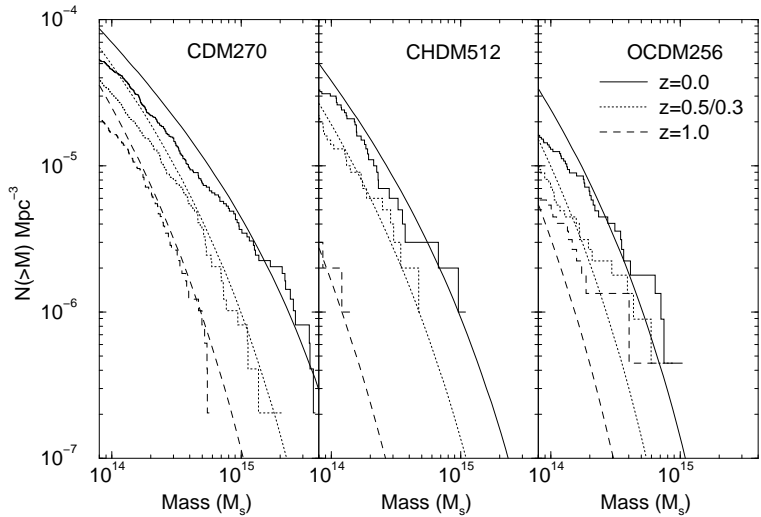}}
\caption{
The mass distributions function for the same models and redshifts as in
Figure~\protect\ref{fig:mass_sigma}.  Lines are Press-Schechter
predictions with $\delta_c$ given by the spherical collapse model (see
text).
}
\label{fig:nm}
\end{figure*}

In Figure~\ref{fig:nm} we show the result of comparing halos from
the simulation (identified with the spherical overdensity scheme
described earlier) with predictions from
equation~(\ref{eq:press_schechter}) for the CDM270, CHDM512 and OCDM256
models at a few different redshifts.  The agreement is generally good,
although there are systematically fewer low mass clusters in the
simulation than predicted.  This is due in part to the fact that those
clusters are poorly resolved, but also seems to be a general
shortcoming of the PS method (\cite{bon96b}).  
All the simulations produced fewer high-mass clusters with increasing
redshift, although the evolution is strongest in the model
with massive neutrinos and weakest in the open cosmology.
The number of massive clusters at higher redshifts (particularly
around $z=1$) for the OCDM256 model is underpredicted.  If this result
is correct, then it makes the difference between open and flat
models even larger than previously predicted by PS methods,
however two comments are in order.  First, this simulation has the
smallest number of well-resolved clusters (CHDM512 has better
resolution, while CDM270 has more and larger clusters).  Second, the
significance is small since the number of clusters involved is very
small.  Indeed, we should point out that these simulations are not the
best to test the Press-Schecter methodology since purely N-body
simulations can afford better resolution, and can use a larger box to
produce more high mass clusters.  Still, we press on to illustrate
what we believe to be a more objective and informative way to
constrain the parameter $\delta_c$.

We can be more quantitative about the fits by varying this parameter
(fixed in Figure~\ref{fig:nm} by the spherical collapse
prediction).  Since the PS formalism is often used
to constrain theories of structure formation, knowledge of the
uncertainty in its predictions is important.  We construct a
likelihood estimator by considering the distribution of clusters 
above a mass cutoff $M_\star$ for each simulation.  If we divide the
range of masses into intervals $\Delta M$ that are sufficiently small
so that the predicted number of objects in each range, $n(M) = \Delta
M dn/dM \ll 1$, then the probability of finding either 0 or 1 objects
in this range is given by the Poisson distribution: $P(0) =
\exp[-n(M)]$ or $P(1) = n(M) \exp[-n(M)]$.
The relative likelihood of finding a particular
set of clusters is the product of the probabilities for all the
mass intervals, split into those with clusters and those without:
$P(\delta_c) = \prod P(1) \prod P(0)$.  For $\Delta M \ll 1$, this
can be written as:
\begin{equation}
P(\delta_c) = \left[ \prod_i\frac{dn(M_i)}{dM} \Delta M \right]
	\exp\left[-\int_{M_\star}^\infty \frac{dn(M)}{dM} dM\right].
\end{equation}
The product is over all the cluster masses $M_i$ above the cutoff
$M_\star$; the distribution is normalized, so that $\int P(\delta_c)
d\delta_c = 1$.  To imitate the simulation as closely as possible, we
have set the power spectrum to zero for scales larger than the
fundamental wavelength of the box.  This has only a slight effect,
shifting $\delta_c$ by about 1\%. 

\begin{figure}
\epsfxsize=3.2in
\centerline{\epsfbox{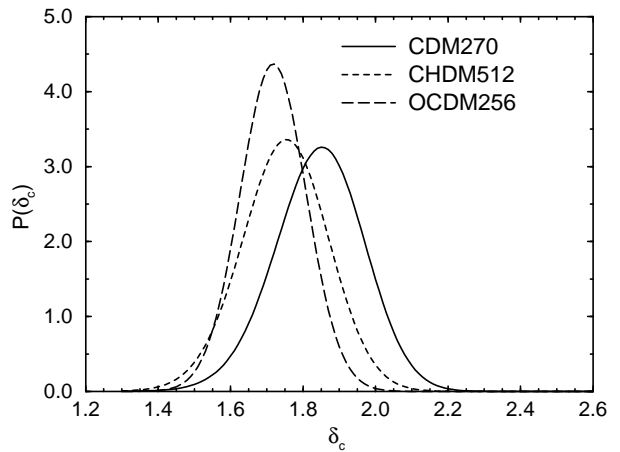}}
\caption{
The likelihood distributions for $\delta_c$, as determined by our three
models at $z=0$.
}
\label{fig:dc_dist}
\end{figure}

Figure~\ref{fig:dc_dist} shows the resulting likelihood distributions
for our three models.  Since we are primarily interested in the higher
mass clusters, we choose a mass cutoff ($M_\star$) of twice the
non-linear mass (this is about $1 \times 10^{14} M_\odot$ for the
OCDM256 and CHDM512 cases and $6 \times 10^{14} M_\odot$ for
CDM270). The curves are reasonably well fit by Gaussians with means
and one sigma uncertainties given by $1.84\pm0.12, 1.75\pm0.12$ and
$1.72\pm0.09$ for CDM270, CHDM512 and OCDM256 respectively.  These are
all compatible with the spherical collapse prediction.

These values are in the range advocated by other authors
(\cite{efs88}; \cite{nar88}; \cite{car89}; \cite{bon91}), although
some caution is in order since we are comparing results with different
methods for identifying clusters and their masses.  For more recent work
which adopts the same methodology as employed here, the agreement is
remarkable.  Lacey \& Cole (1996)\markcite{lac96} used scale-free
simulations with a variety of cluster analysis techniques.  For the
spherical overdensity method with a top-hot filter, they find a best
fit of $\delta_c = 1.96$, however, this is over a considerable range
in mass. For their rarest peaks (corresponding to the clusters
discussed here), it appears that a normalization somewhat lower, but
just above 1.68, is appropriate.  Eke, Cole \& Frenk
(1996)\markcite{eke96} found that a value around 1.75 gave good fits
to very large ($350^3 h^{-1}$Mpc$^3$) simulations for a variety of
cosmologies.  Borgani \etal (1997)\markcite{bor97} argue for
$\delta_c$ near 1.68 (for a top-hat filter) based on a series of
simulations including a cosmological constant term.

Finally, we note that a cluster identification based on mean
overdensity does not necessarily produce a one-to-one correlation with
bound halos.  As we discuss in the section on temperature, this should
not strongly affect our results, however it is important if one is
interested in the details of bound halos.  For a
more complete discussion of the mass function of halos in the CDM
model see Gelb \& Bertschinger (1994)\markcite{gel94}.

\begin{figure}
\epsfxsize=3.1in
\centerline{\epsfbox{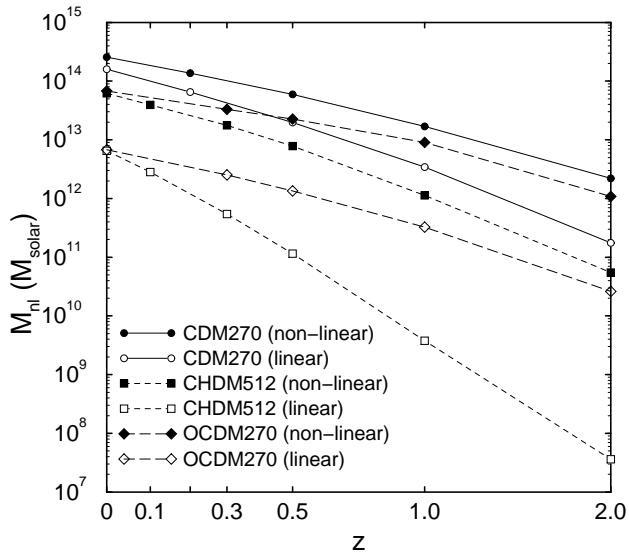}}
\caption{
The evolution of the non-linear mass (see text for definition) for our
three primary models, computed both through the linear and
non-linear (N-body) power spectra.  
}
\label{fig:mass_nl}
\end{figure}

The high-end of the mass function is exponentially sensitive to the
amplitude of the power spectrum, so it is worthwhile examining the
non-linear mass ($M_{nl}$), which controls the position of this
cutoff.  It is defined as the mass within a spherical
volume for which the rms overdensity from equation~(\ref{eq:sigma}) is
equal to $\delta_c$.  Figure~\ref{fig:mass_nl} shows the evolution of this
quantity, using both the linear power spectrum and the fully
non-linear (N-body) spectrum (for a more complete discussion in terms
of the CDM model, see \cite{jai94}).  This figures shows both
the increased normalization of the CDM270 model, and the slower
evolution of the OCDM256 mass function.  Scale-free spectra with
$P(k) \propto k^n$ result in $M_{nl} \propto (1+z)^{-6/(n+3)}$.
The ordinate is $\log(1+z)$ so the these models would show up as
straight lines on the plot.  The linear results for CDM are
closer to the non-linear than for the other models because of the
shallower spectral profile (smaller logarithmic slope).  In other
words, the CHDM slope is close enough to $P(k) \sim k^{-3}$ that the
extremely rapid evolution of $M_{nl}$ requires more accuracy to obtain
a good estimate of the non-linear mass.  

\begin{figure*}
\epsfxsize=4in
\centerline{\epsfbox{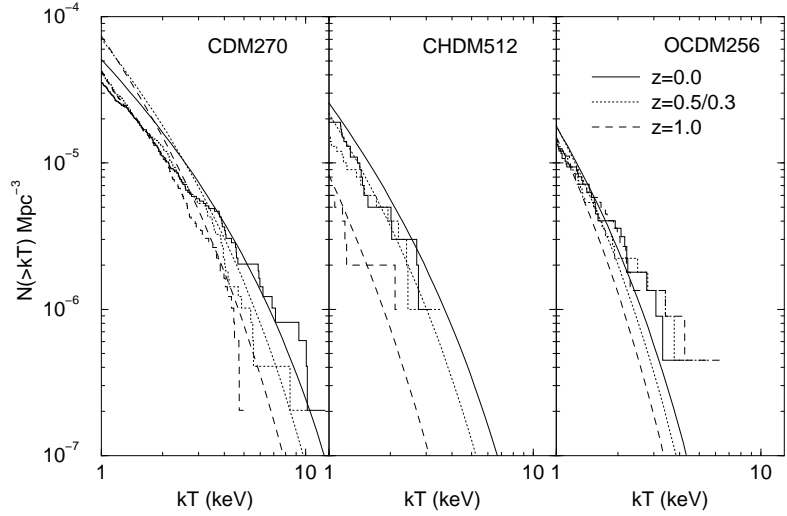}}
\caption{
The temperature distribution functions for the same models and
redshifts as in Figure~\protect\ref{fig:mass_sigma}.  Lines are
predictions from the extended PS model with $\delta_c$ given by the
spherical collapse model and $f_T = 0.77$.
}
\label{fig:nt}
\end{figure*}

Although useful, the mass distribution is difficult to obtain from
observations because of uncertainties in obtaining the total mass out
to the virial radius.  Somewhat easier is the temperature distribution.
In Figure~\ref{fig:nt}, we plot this function for the canonical
models as well as analytic predictions using Press-Schechter and the
mass-temperature relation: equations~(\ref{eq:press_schechter}) and
(\ref{eq:temp_mass}).  From the results of section~\ref{sec:scaling},
we adopt the mass-temperature normalization $f_T = 0.77$, producing
reasonable agreement between the simulations and analytic results.  
Again, there is are strong difference between the evolutionary
properties.  When medium to high redshift temperature samples
become available, they will provide strong constraints. 

We stress at this point that the temperature (and velocity dispersion)
distribution functions are much more robust against variations in
cluster identification shemes than the mass function.  This is true
because the mass is a rising function of radius (since $\rho \sim
r^{-2}$), while, to first order, the temperature is constant.
How can $N(>T)$ can be relatively independent of the cluster
identification method, while $N(>M)$ --- from which it is derived ---
is not?  In fact, this is only true if the $N(>M)$ and the $T$-$M$
relation are calibrated {\em consistently}.  In other words, if we
change the way we find the cluster mass (within reason), the $T$-$M$
relation and the mass function will both change such that the
resulting $N(>T)$ is unchanged.   This feature of the
temperature distribution is particularly important when it comes time
to compare with observations, which are rarely determined to the
virial radius.

\begin{figure*}
\epsfxsize=4.2in
\centerline{\epsfbox{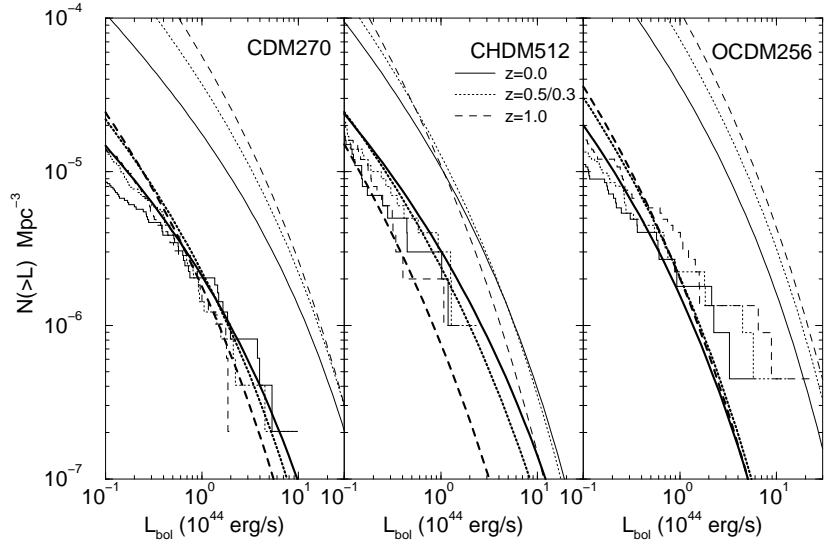}}
\caption{
The bolometric luminosity distribution functions for the same models
and redshifts as in Figure~\protect\ref{fig:mass_sigma}.  The
Press-Schechter results for two mass-luminosity relations are also
shown: thin lines use equation~(\protect\ref{eq:mass_lumin_nfw}) while
bold lines include the effects of limited resolution:
equation~(\protect\ref{eq:mass_lumin_code}).
}
\label{fig:nl_bol}
\end{figure*}

Since luminosities are more easily obtained observationally, we use
equation~(\ref{eq:mass_lumin_nfw}) to determine the bolometric
luminosity distribution with Bremsstrahlung radiation and zero
metallicity (this constraint will be relaxed in the next section).
These are shown as thin lines in Figure~\ref{fig:nl_bol}.  While the
simulated clusters do not match these results, including the effects
of finite resolution with equation~(\ref{eq:mass_lumin_nfw}) does
produce agreement, at least for the relative bright clusters.
The `fully' resolved Press-Schechter predictions show that, for the
brightest clusters, there is positive evolution with redshift in the
OCDM model but negative evolution for CHDM.  However, the exact
behaviour depends on the adopted $L$-$T$ relation.  Since the fixed
resolution produces an $L-T$ relation closer to that observed (albeit
for numerical reasons), it may be the better predictor of evolution.
In that case, the OCDM model is nearly constant over redshift, while
the CHDM evolution is strong negative.

Perhaps more importantly, the difference between the thin and bold
lines indicates the kind of error engendered by our limited spatial
resolution.  Roughly speaking, there are two effects.  The first is
the obvious reduction in the luminosity of all clusters, shifting the
curves to the left.  The other is to add an additional component of
negative redshift evolution, exaggerating the negative evolution at
high luminosities and decreasing the positive evolution at the low
end, causing the curves to tilt slightly.  This second
effect is relatively small, about the size of Poisson fluctuations due
to the limited number of clusters.  

\subsection{Metallicity Effects}
\label{sec:metallicity}

In this section we address more realistic X-ray emissivity effects:
limited band-passes and line emission from multiply ionized heavy
elements in the plasma: metallicity.  As we will show, for moderate
and high temperature clusters, the ratio of line to continuum emission
is actually quite small, only a few percent, making the accurate
determination of a cluster's metallicity difficult.  On the positive
side, this make our results largely insensitive to the assumed
metallicity.  Nevertheless, the high resolution spectroscopy performed
by the ASCA satellite, as well as work done with ROSAT and other
missions, seems to indicate that a value of $Z/Z_\odot = 0.3$ in solar
units is reasonable for many clusters (\cite{kow93}; \cite{mar94}).

\begin{figure}
\epsfxsize=3.1in
\centerline{\epsfbox{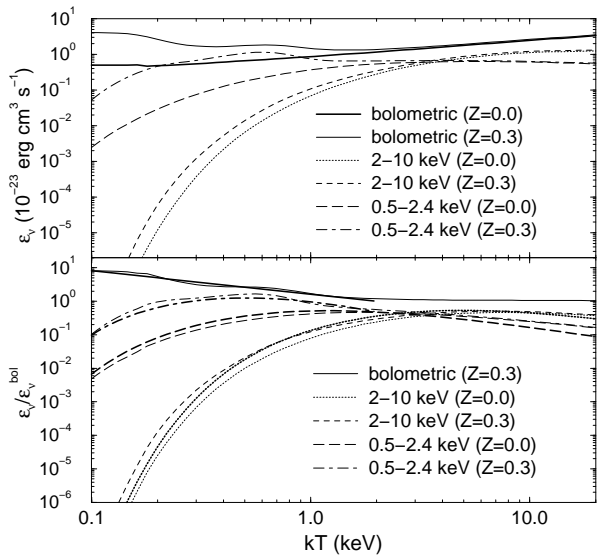}}
\caption{The X-ray emissivity as a function of temperature
for two different metallicities and three different band-passes from a
Raymond-Smith code
(top panel).  The relative X-ray emissivity (normalized
to the bolometric emissivity at that temperature for a metal-free
plasma) as thin lines (bottom panel).  Also shown in thick lines are
the approximate fits from equation~(\protect\ref{eq:metallicity_fit}). 
}
\label{fig:metallicity}
\end{figure}

We use a Raymond-Smith code (1977, 1992 version)\markcite{ray77} to
compute emissivities which are shown in the top panel of
Figure~\ref{fig:metallicity} for a variety of metallicities and
band-passes as a function of temperature assuming a number density of
electrons of $10^{-3}$ cm$^{-3}$.  The effect of line emission from
metals is small for $T > 2$ keV.  This means that previous results
(\cite{cdm_xray},b) computed with the assumption of zero-metallicity
will not be substantially affected in the region where observational
results are most plentiful.  Nevertheless, this will prove important
for small clusters and groups of galaxies so we extend our previous
analytic results to include this effect (although, of course, our
simulations do not include many other effects that may prove important
for small clusters such as stellar feedback and radiative cooling).

To do this, we fit simple expressions to the full result by ignoring
the temperature variation of the gaunt factor and approximating the
effect of metallicity as a power law with a cutoff.  The resulting
formula is given in relation to the bolometric luminosity ($L_{metal}
= \epsilon(Z, T, \nu_1, \nu_2) L_{bol}$, where $\nu_1$ and $\nu_2$
specify the band-pass) by:  
\begin{eqnarray}
\epsilon(Z, T, \nu_1, \nu_2) &=& 
     \left[\exp(\frac{-h\nu_1}{kT}) - \exp(\frac{-h\nu_2}{kT}) \right]
     \times \nonumber \\  & & \hspace*{-1.4cm}
     \left\{\begin{array}{ll}
	   (kT/2\hbox{ keV})^{-\gamma\sqrt{\frac{Z}{0.3 Z_\odot}}} &
			kT < 2 \hbox{ keV} \\
		1 &	kT > 2 \hbox{ keV} 
		\end{array} 
	\right.
\label{eq:metallicity_fit}
\end{eqnarray}
The effect of metallicity depends on the bandpass used and cannot be
simply modelled, so we parameterize it through $\gamma$ and
tabulate values appropriate for a few common choices: $\gamma = 0.6$
for 0.1--2.4 keV; 0.9 for 0.5--2.4 keV; 0.0 for 2--10 keV; and 0.7 for
bolometric luminosities (we use $g_{ff}=1.3$, and 24\% He by mass).
This fit is accurate to only about a factor of two, but will be
sufficient for our purposes; a few examples are shown in the bottom
panel of Figure~\ref{fig:metallicity}. 

\begin{figure}
\epsfxsize=3in
\centerline{\epsfbox{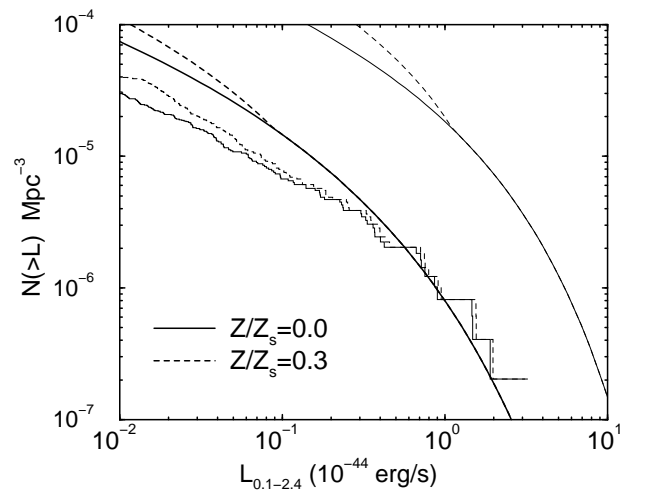}}
\caption{The 0.1-2.4 keV luminosity distribution
function with (dashed) and without (solid) metallicity for the CDM270
model. The extended Press-Schechter results are also shown, with the
usual meaning for thin and thick lines.
}
\label{fig:nl_0.1_2.4}
\end{figure}

With this extension we can now compute more observationally oriented
luminosity functions and compare them to the luminosity function
computed directly from the simulation.
The numerical cluster luminosities are
computed on a cell-by-cell basis so they include the spatial variation
of temperatures (see section~\ref{sec:profiles}), which the analytic
models are missing.  They also use emissivities calculated from the
full Raymond-Smith code.  Figure~\ref{fig:nl_0.1_2.4} shows the
luminosity function in the 0.1-2.4 keV bandpass for a representative
case (CDM270 at $z=0$).  The PS model overpredicts the number of
low-luminosity clusters, just as it does for low mass clusters,
however the general effect of metallicity is included correctly.
This also demonstrates that the effects of metallicity are negligible
above about $10^{44}$ erg/s and relatively slight above $10^{43}$
erg/s.

\section{Cluster structure}
\label{sec:profiles}

In deriving some of the analytic results we assumed a spherically
symmetric isothermal profile for both the gas and dark matter.  In
this section we examine the simulated clusters in order to determine
the accuracy of these assumptions.  A more complete analysis will be
presented in Bryan \& Norman (1997b)\markcite{bry97b}.

\begin{figure}
\epsfxsize=3.1in
\centerline{\epsfbox{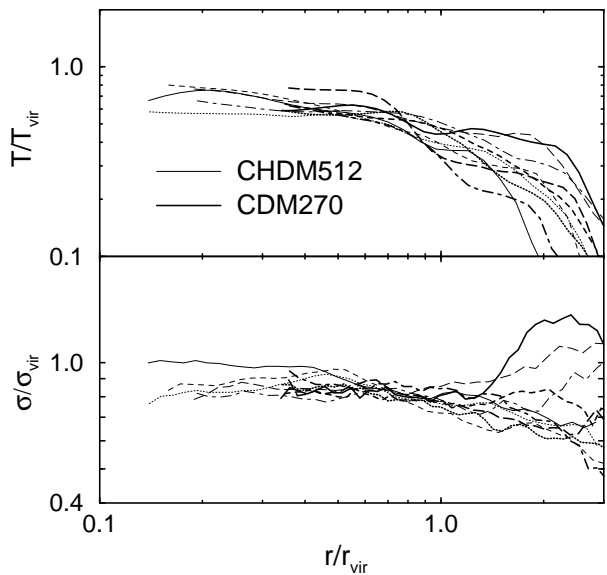}}
\caption{
Temperature (top panel) and dark matter one-dimensional velocity
dispersion (bottom panel) profiles for the CDM270 and
 the CHDM512 models at $z=0$, normalized by their respective
virial values.
}
\label{fig:radial_profiles}
\end{figure}

In Figure~\ref{fig:radial_profiles} we show profiles of 
temperature and the one-dimensional velocity dispersion for the five
most massive clusters in our two canonical models.  These are
normalized by their appropriate virial values (with $f_T = f_\sigma =
1$).  To compute the profile, we redetermine the cluster centers
by adopting the center of the cell with the highest gas density within
$r_{vir}/2$ of the original center found through the iterated
spherical overdensity method.  This procedure does a good job of
finding the core of the largest mass clump.  The innermost point
plotted is two cell widths from the cluster center.  This is
approximately our resolution limit.

The spherically-averaged temperature profiles are compatible between
models and show a very slowly falling profile to about $0.7 r_{vir}$,
a somewhat steeper slope ($r^{-0.5}$) to about two times the virial
radius and then a very sharp fall off beyond that.  
The velocity profiles are close to flat, although they appear to show
a slight dip at and just before the virial radius.  They are also
slightly below their respective $\sigma_{vir}$ values (implying
$f_\sigma < 1$).  However, this tendency may be declining
as we resolve further towards the center.
The hot component (we plot only the cold particles
in Figure~\ref{fig:radial_profiles}) follows almost exactly the same
profile at large r, but is systematically higher at low radii.
The temperature profile shows much
greater variation and is less compatible with an isothermal model, even after
spherical averaging.  The temperature and $\sigma$ profiles are in
agreement with the factors $f_T\sim 0.8$ and $f_\sigma \sim 0.85$-0.9
derived earlier.


\section{Conclusion}
\label{sec:chap6_conclusion}

In this paper, we have shown that X-ray clusters produced in
Eulerian simulations agree well with the scaling relations involving
mass, temperature and the collisionless velocity dispersion.  The
luminosity behaviour is more affected by resolution but can be simply
and accurately modelled.
The predicted bolometric $L$-$T$ relation does not match that
observed.  We argue that this is mostly the fault of the luminosity
prediction since it is very sensitive to the structure of the
clusters, while the temperature is not.  Indeed, the mass-temperature
relation appears to be remarkably robust and, as we have demonstrated,
depends very little on the input physics, numerical method,
resolution, or cosmology.  Our result, combined with a review of the
available literature, indicates that $f_T \approx 0.8$ in the notation
of equation~(\ref{eq:temp_mass}).
We also demonstrated that the isothermal profile assumed in computing
the normalization of the scaling relation between $M$ and
$\sigma$ is a reasonable approximation for the dark matter (although
there is evidence for $f_\sigma < 1$).

We stress here the difference between $f_T$, $f_\sigma$ and $\beta$.
The first two adjust the virial scaling relation normalizations as
compared to the hydrostatic isothermal sphere assumption, while
$\beta$ is the ratio of the collisionless `temperature' to the gas
temperature, equation~(\ref{eq:chapter6_beta}).  These three
quantities are related through $\beta = f_\sigma^2/f_T$.  Since
prescriptions such as the Press-Schechter formalism predict the
distribution of masses, the important normalizations are $f_\sigma$
and $f_T$ (of the $M$-$\sigma$ and $M$-$T$ relations, respectively),
and not $\beta$, as is occasionally assumed (e.g. \cite{eke96}).

\begin{figure}
\epsfxsize=3.1in
\centerline{\epsfbox{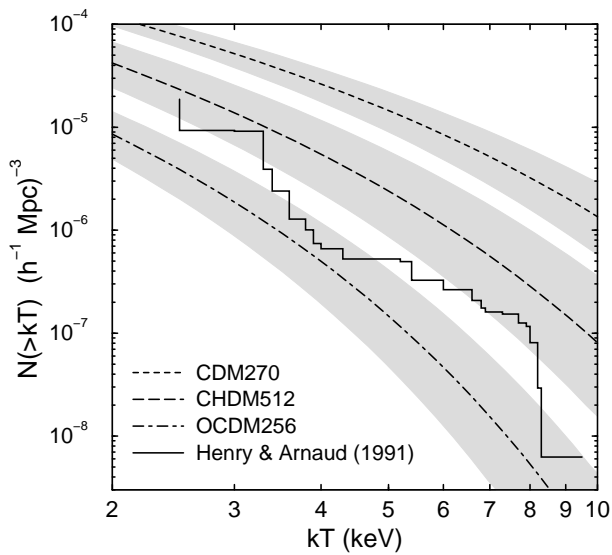}}
\caption{
The PS temperature distribution functions for our three primary models at
$z=0$ using the best fit $\delta_c$ for each model.  Shaded
regions are the two-sigma uncertainties in $\delta_c$.  Observations
from Henry \& Arnaud (1991)\protect\markcite{hen91} are shown as a solid line.
}
\label{fig:nt_observe}
\end{figure}

Based on these findings and the Press-Schechter prescription for
computing the differential number density of virialized halos, we
computed the temperature distributions functions which are in good
agreement with those derived from the numerical simulations (over
their range of validity).  In Figure~\ref{fig:nt_observe}, we show the
temperature functions for our three models along with the 95\%
uncertainty in $\delta_c$.  The observations (\cite{hen91};
\cite{eke96}) are also shown, computed as $N(>T) = \sum_{T_i>T}
1/V_{max,i}$ (where $V_{max,i}$ is the maximum volume for which the
cluster with temperature $T_i$ and flux $f_i$ could have been
observed).  This shows that the CDM270 model predicts substantially
too many clusters, while 
the CHDM512 and OCDM256 models are (very) marginally in agreement.
The fits would be much improved if the power spectra were renormalized
to $\sigma_8 = 0.55, 0.63, 0.85$ for CDM270, CHDM512 \& OCDM256.  To
retain agreement with COBE, the CHDM model would require only a slight
tilt.
This result differs slightly from Ma (1996)\markcite{ma96} due to the
addition of a second neutrino.  Otherwise, for the same $Q_{rms-PS}$
we would have to increase $\sigma_8$ by 10\% (\cite{sto95}).
The OCDM270 model could also be adjusted slightly to match both COBE
and clusters, however, CDM would require a large tilt in the spectrum
($n \sim 0.8$).

Analytic luminosity functions computed with the effects of finite
resolution (Figure~\ref{fig:nl_bol}) agree well with the simulations
and allow us to gauge the impact of resolution as a function of
luminosity and redshift.  An extension which accounts for metallicity
and limited bandpass, equation~(\ref{eq:metallicity_fit}), can be
used to compare directly against observational data.
Just as we used the temperature distribution function, we could also
use the luminosity to compare to observations, however, the
observational luminosity-temperature relation does not agree with the
scaling relations.  Using the observed $L$-$T$ relation to compute
$n(>L)$ from $n(>T)$ does not add any additional information, since if
we match the temperature function we must also match the luminosity
function.  However, recent observations with the ASCA satellite have
produced a sample of temperatures at moderate redshift (\cite{mus97}).
The sample is not complete so we cannot use it to construct a
temperature function, however it can be used to constrain evolution in
the observed $L$-$T$ relation.  Combined with the method for
predicting $n(>T)$ described in this paper and the upcoming high
redshift luminosity samples (\cite{rom97}; \cite{sch97}), we will be
able to place much tighter constraints on the models based on cluster
evolution.




\vspace{0.2cm}
This work is done under the auspices of the Grand
Challenge Cosmology Consortium and supported in part by NSF grants
ASC-9318185 and NASA Long Term Astrophysics grant NAGW-3152.

\end{document}